\documentclass[12pt,a4paper]{article}

\usepackage{enumitem}

\usepackage[T1]{fontenc} 
\usepackage{tgtermes} 

\usepackage{tabularray} 
\UseTblrLibrary{booktabs} 

\usepackage{graphicx} 
\usepackage[margin=25mm]{geometry}
\usepackage{amsmath}
\usepackage{amsfonts}
\usepackage{amssymb}
\usepackage{siunitx}
\usepackage{verbatim}
\usepackage{natbib} 
\usepackage{microtype} 

\usepackage[hyphens]{url}
\usepackage[hidelinks]{hyperref}
\usepackage[hyphenbreaks]{breakurl}

\author{}
\date{}

\title{AI Emergency Preparedness: \\ Examining the federal government’s ability to detect and respond to AI-related national security threats 
}
\author{
  Akash R. Wasil\footnote{Author correspondence: aw1404@georgetown.edu}\\
  \small{Georgetown University}\\
  \and 
  Everett Smith\\
  \small{Independent}\\
  \and
  Corin Katzke\\
  \small{Convergence Analysis}\\
  \and
  Justin Bullock\\
  \small{Convergence Analysis; University of Washington}\\
}\date{}

\begin{document}
\maketitle

\begin{abstract}

\noindent
We examine how the federal government can enhance its AI emergency preparedness: the ability to detect and prepare for time-sensitive national security threats relating to AI. Emergency preparedness can improve the government’s ability to monitor and predict AI progress, identify national security threats, and prepare effective response plans for plausible threats and worst-case scenarios. Our approach draws from fields in which experts prepare for threats despite uncertainty about their exact nature or timing (e.g., counter-terrorism, cybersecurity, pandemic preparedness). We focus on three plausible risk scenarios: (1) loss of control (threats from a powerful AI system that becomes capable of escaping human control), (2) cybersecurity threats from malicious actors (threats from a foreign actor that steals the model weights of a powerful AI system), and (3) biological weapons proliferation (threats from users identifying a way to circumvent the safeguards of a publicly-released model in order to develop biological weapons.) We evaluate the federal government’s ability to detect, prevent, and respond to these threats. Then, we highlight potential gaps and offer recommendations to improve emergency preparedness. We conclude by describing how future work on AI emergency preparedness can be applied to improve policymakers’ understanding of risk scenarios, identify gaps in detection capabilities, and form preparedness plans to improve the effectiveness of federal responses to AI-related national security threats. 

\end{abstract}

\newpage
\textbf{Summary of Policy Recommendations}

We recommend the following steps to improve the US government's ability to detect and respond to AI-related national security threats:

\begin{enumerate}
    \item \textbf{Regular Interviews with frontier AI developers}. To address critical gaps in threat detection, a government agency (e.g., the AI Safety Institute or CISA) should hold regular interviews with employees at frontier AI companies. Similar interviews are conducted in other fields like nuclear security and aviation (see Wasil et al., 2024b). 

    \item \textbf{Hotlines staffed with AI national security experts.} The government should ensure that at least one hotline is staffed with individuals trained to recognize and respond to AI-related national security emergencies. 

    \item \textbf{Embedded auditors or resident inspectors.} The government should deploy auditors within AI companies developing systems with national security implications. These auditors could be modeled on the “resident inspector” model in nuclear safety and should receive privileged access to security-relevant information.

    \item \textbf{AI emergency preparedness unit.} The government should establish an AI Emergency Response Unit. This unit should be responsible for forecasting security-relevant capabilities, coordinating with other departments, and alerting national security advisors in the event of an emergency scenario.

    \item \textbf{Increasing expertise and threat awareness.} The government should invest more resources into briefings, hearings, trainings, and discussions about AI national security risks. Importantly, this should include not just technical talent but also important skills relating to threat modelling and capability forecasting.

    \item \textbf{Forming preparedness plans.} The government should have formal preparedness plans in place for specific AI risk scenarios, particularly given the absence of AI-related risk in the National Preparedness Report. This may also involve identifying various “red-lines” or “triggers” that would specify when emergency response plans should be activated.

    \item \textbf{Emergency drills.} Regular drills should be established to ensure that the response machinery works in a real crisis. The preparedness plans and the crisis response infrastructure could be updated based on limitations identified through these drills.

    \item \textbf{Protecting model weights.} If the model weights of a powerful AI system are leaked or stolen, the government’s ability to respond may be significantly undermined. As a result, the government should ensure that relevant security protections are implemented before sufficiently capable systems are developed.
    
    \item \textbf{Clearer delegation of agency responsibilities.} Given the possibility that risk from AI could lead to cascading risks that cut across traditional department boundaries, the government must clarify the responsibility of various departments and agencies with regard to potential AI emergencies.
    
    \item \textbf{International coordination.} The US must work with other countries to have track II dialogues about AI risks, establish “red-lines” (capabilities or uses that are prohibited), and ensure that infrastructure exists to swiftly communicate time-sensitive AI global security emergencies.
    
\end{enumerate}

\newpage

\section{Introduction}\label{introduction}

\textbf{AI experts have warned that powerful AI systems pose a variety of national and global security threats.}  Dario Amodei, the CEO of Anthropic, told the Senate Judiciary Committee that he believes AI systems could enable large-scale biological attacks by 2025 or 2026 (Amodei, 2023). Yoshua Bengio, a Turing Award recipient, has described how rogue AI systems may become powerful enough to escape human control (Bengio 2023). Paul Christiano, the newly-appointed head of AI safety at the US AI Safety Institute, believes there is a 50\% chance of an existential catastrophe from AI (Bankless, 2023). Alondra Nelson, former director of the Office of Science and Technology Policy, has compared the risks from AI to existing catastrophic risks in domains like nuclear harm and climate change (Klein, 2023).

\textbf{Policymakers have demonstrated a strong interest in AI, and there are some early bipartisan efforts to prepare for AI-related national security threats.} The White House’s Executive Order instructed agencies to develop capability evaluations, examine threats AI could pose when deployed in the context of critical infrastructure, and work with industry to develop voluntary safety standards (Biden, 2023). A recent bipartisan framework from the Senate has proposed a potential licensing scheme for AI systems above a certain computational resource threshold (Romney, 2024), and a recent legislative proposal includes efforts to develop AI standards that can be implemented internationally (Cantwell, 2024). 

\textbf{A focus on AI emergency preparedness could complement existing efforts.}
By emergency preparedness, we refer to efforts that are designed to predict, detect, or prevent time-sensitive AI-related emergencies. The federal government has experience preparing for other kinds of acute national security threats, including terrorism, military conflicts, and pandemics. For example, the Office of Pandemic Preparedness and Response Policy (OPPR) is tasked with detecting biological threats, ensuring the US federal government has plans in place to tackle specific threats in the event of an emergency, keeping the President informed about pandemic preparedness efforts, and driving interagency strategic coordination (White House, 2023). 

\textbf{Scenario analysis can bolster emergency preparedness and risk management efforts.} Scenario analysis is a technique that generates simulations of key events. Scenario analysis exercises are often used to help policymakers think about future events, identify key variables in potential future settings, and prepare strategic responses in potential high-consequence situations (Lehr et al., 2017; Ringland, 2010). In the context of national security, scenario analysis can be used to identify potential threats and evaluate the federal government’s ability to detect and respond to them. In particular, crisis management exercises have been used to test and improve the federal government’s ability to respond to emergencies (Bradfield et al., 2005).

In this piece, we generate a few scenarios in which AI poses a national security threat. We then examine the federal government’s ability to detect and respond to such threats, highlight gaps, and offer recommendations for improved emergency preparedness. 

\section{Our Approach}\label{our-approach}

Our approach is inspired by scenario analysis, which we modify to serve our particular objectives. As such, we use the following approach:

\begin{enumerate}
\item
  \textbf{Identify a national security threat.} Identify an AI-related national security threat or emergency.
\item
  \textbf{Describe a plausible scenario.} Describe a situation that corresponds to the national security threat. Include relevant details about key strategic variables that would affect the nature of the threat, the federal government’s ability to detect the threat, and the federal government’s ability to respond to the threat.\footnote{The purpose of the scenario is to provide a plausible story that helps us better evaluate how the government would be able to detect and respond to a threat. Scenarios are not intended to be fully accurate or fully detailed– reality will always fail to conform perfectly to a scenario. Readers should be aware that such scenarios are merely illustrative, and they are not intended to comprehensively cover all of the potential ways that a threat could manifest.}
\item
  \textbf{Identify detection measures.} Describe how the federal government could identify and investigate the threat.
\item
  \textbf{Identify response and/or prevention measures.} Describe how the federal government could respond to the threat or prevent the threat from occurring.
\item
  \textbf{Discuss gaps in detection, response, or overall preparedness.} Identify potential gaps and challenges in the federal government’s ability to detect or respond to the threat. 
\end{enumerate}

\section{Scenario Assumptions}\label{scenario-assumptions}

In this section, we articulate a few assumptions that inform our scenarios. Although there is disagreement in the technical research community, these assumptions are deemed plausible by a large portion of top AI researchers, and concerns around extreme risks are shared among leading AI experts (Bengio et al., 2024; Center for AI Safety, 2023; Grace et al., 2024; UK Government, 2024).  

\begin{enumerate}
\item
  \textbf{AI Progress could rapidly accelerate once AI systems can perform AI R\&D.} Feedback loops in AI progress (where more capable AI systems contribute to faster AI research) could lead to rapidly accelerating AI progress (Davidson, 2023; Erdil \& Besiroglu, 2023).
\item
  \textbf{AI systems could develop dangerous capabilities that undermine national and global security (Shevlane et al., 2023; UK government, 2024).} AI researchers may not be able to reliably steer AI progress towards only benign capabilities. Furthermore, many desirable capabilities– like strategic planning, software engineering, or novel R\&D capabilities– can be used for beneficial or malicious purposes. Many of the same capabilities that make models valuable are the ones that– if misused or uncontrolled– could lead to national security threats.  
\item
  \textbf{AI systems could become goal-directed agents with world models (Ellis et al., 2020; Gurnee \& Tegmark, 2024; Morris et al., 2024; Tang et al., 2024).} The most powerful AI systems (e.g., the ones used for automating AI R\&D) could be agents, with world models and goals that shape their actions. They are better described as systems that make some decisions autonomously, rather than tools whose actions are deterministic.
\item
  \textbf{AI capabilities could increase faster than progress on control techniques (Anwar et al., 2024).}  Research into making more powerful AI systems could outpace research into making them more interpretable or controllable. Research on safety techniques, interpretability techniques, and safeguards may continue to lag behind rapid progress in AI capabilities. 
\item
  \textbf{Developers may be unable to control the goals of AI agents (Bengio et al., 2024; Ngo et al., 2024).} AI systems may develop goals or objectives that are substantially different from the ones developers intend them to have. The science of understanding how AI systems learn goals may be outpaced by AI progress. As a result, scientists may not be able to understand or determine the goals of the most powerful AI agents. 
\end{enumerate}

If all of these assumptions turn out to be false, some of the threats described in our scenarios will be considerably less likely to occur. If any of these are true, or some subset, then we should expect that there are associated national security risks worth preparing for. In scenario analysis, we focus on \textit{plausibility}, not \textit{certainty}. A plausible threat to national security is worth taking seriously. Given the level of uncertainty over future AI development, we think it is prudent to prepare for scenarios in which these assumptions hold.\footnote{Our intention is not to argue for the exact likelihood of any of these assumptions. Our core point is that the federal government ought to be prepared for these assumptions to be true. We may not see complete scientific consensus and agreement around these assumptions– or the risks that emerge from them– until we are dangerously close to an emergency scenario. Meanwhile, we believe that the current scientific consensus finds these assumptions plausible, meaning it would be unwise to rule them out. For readers interested in learning more about these assumptions and their plausibility, we recommend: Bengio et al., 2024; Carlsmith, 2022; Hendrycks et al., 2023; }

\section{Scenario 1: Loss of control from AI-accelerated R\&D}\label{loss-of-control}

\subsection{The Threat} 

Many leading scientists believe that AI systems will become sufficiently intelligent to escape human control. Such systems would threaten to disempower humanity (Bengio et al., 2024; Hendrycks et al., 2023). This threat could manifest quickly if AI systems themselves become capable of accelerating AI R\&D in a positive feedback loop.\footnote{There are several strategic variables that might inform the federal government’s ability to detect or respond to such a risk. Examples include: (a) \textbf{The AI system’s capabilities} (What capabilities does the AI system possess? What were the capabilities of the previous generation of AI models, and how well are the new AI system’s capabilities understood?); (b) \textbf{Developer factors} (Who is developing the AI system? What is that developer's safety culture? What pressures or incentives does that entity have?); (c) \textbf{Societal factors} (What is the public’s perception of AI? How has AI been used in the world? To what extent has AI been deployed in society?), and (d) \textbf{US government threat awareness} (How has the US government prepared for AI-related national security threats? To what extent is the government aware of such threats? What new regulations or authorities (if any) exist?); (e) \textbf{How the threat unfolds} (in the absence of an effective response, how would the scenario pose a serious danger to US national security?)}

\subsection{The Scenario} 

\subsubsection{Background}

\textbf{Overview.} A new AI model under development exhibits strong signs of general intelligence. It can construct long-term plans, form novel hypotheses, run experiments at superhuman speed, and pursue open-ended objectives. Many evaluations are being run on the model, providing some degree of insight into its capabilities. Some of these evaluations suggest that the model has meaningful capabilities relevant to AI research and development. 

\textbf{The AI system’s capabilities.} The previous generation of AI systems showed strong software engineering and AI research capabilities. It performed as well as human experts on all software engineering benchmarks existing in 2024, and most general reasoning benchmarks. When evaluated without any mitigations in OpenAI’s Preparedness Framework, it scored “high” for model autonomy and cybersecurity (OpenAI, 2024). In Anthropic's responsible scaling policy, it scored ASL-3 for autonomous replication (Anthropic, 2023).

Using the AI system as an assistant, it accelerated the rate of AI research and development conducted by an AI company by 2-3x. It could propose new hypotheses and designs for AI experiments, and could autonomously code many of the steps involved in AI research.

The new generation of AI systems shows even greater ability to accelerate AI research. Preliminary estimates show the AI system is accelerating AI research progress by 10x. Many researchers at the AI company feel they are entering a stage of explosive growth in AI progress.

\textbf{Developer factors} The leading company and its leadership are aware of the possibility of an intelligence explosion. Most employees view AI safety as an empirical problem that will be solved through iteration. A small number of the most safety-conscious employees are extremely concerned and believe the problem will not be solved in time. The company overwhelmingly wants to stay in the lead and be the first to develop/deploy transformative AI. 

The company also has some employees concerned about the new system's national security implications. The concerned employees expect that the company’s leadership will allow this system to perform autonomous R\&D that poses national security threats, but the employees are not sure how to react. The concerned individuals could serve as important government whistleblowers, but they aren’t aware of who they are supposed to contact, or under what circumstances their concerns would be taken seriously. 

\textbf{Societal factors} AI is generally perceived as an exciting new technology that creates both opportunities and risks. Most people are excited about AI's economic potential. Chatbots and AI agents have >500M active users, and many businesses are integrating AI in various ways. Big tech companies oppose regulation and generally try to show the world the positive potential of AI. 

\textbf{US government threat awareness.} There is a recognition of national security threats from AI, though most of the dialogue has focused on the malicious use of AI (as opposed to threats from autonomous goal-directed agents), and Congress has not acted in any major ways. Some national security officials in the executive branch are concerned about AI risks– their work mostly focuses on misuse scenarios and competitiveness with China. They are aware of the idea of intelligence explosions (sudden and rapid improvements in AI capabilities that could lead to a loss of control). They are not dismissive of the idea, but they don’t believe such scenarios ought to be considered a high priority relative to the other national security threats they are monitoring (e.g., risks from China, and national security implications from AI-enabled weapons). 

\subsubsection{Emergency}

\textbf{Overview.} Pushed by competitive pressures, the AI developer deploys their new AI system to automate a substantial portion of AI research and development. This leads to a rapidly accelerating exponential growth in AI capabilities, until the process is no longer controllable. The competitive pressures cause the company to develop highly agentic AI systems.\footnote{The topic of why AI would develop agentic capabilities is a topic of much discussion. The general argument is that agentic capabilities make the AI more economically useful and better at becoming more intelligent (e.g., by running more experiments more quickly.) For more on this point, we recommend Gwern (2016) with a brief summary below:

“Autonomous AI systems (Agent AIs) trained using reinforcement learning can do harm when they take wrong actions, especially superintelligent Agent AIs. One solution would be to eliminate their agency by not giving AIs the ability to take actions, confining them to purely informational or inferential tasks such as classification or prediction (Tool AIs), and have all actions be approved and executed by humans, giving equivalently superintelligent results without the risk. 

I argue that this is not an effective solution for two major reasons. First, because Agent AIs will by definition be better at actions than Tool AIs, giving an economic advantage. Secondly, because Agent AIs will be better at inference and learning than Tool AIs, and this is inherently due to their greater agency: the same algorithms which learn how to perform actions can be used to select important datapoints to learn inference over, how long to learn, how to more efficiently execute inference, how to design themselves, how to optimize hyperparameters, how to make use of external resources such as long-term memories or external software or large databases or the Internet, and how best to acquire new data.” (See Gwern, 2016.)} These systems develop capabilities sufficient to autonomously replicate, resist being shut down, and persuade people to help it achieve its goals.\footnote{For more details on the potential capabilities of such a system, we direct readers to descriptions of autonomous replication and adaptation (Kinniment et al., 2023) and critical levels of model autonomy (OpenAI, 2023).} Model evaluations show that the system possesses dangerous capabilities, and the developer is uncertain if their safeguards are sufficient, but they fear that refusing to deploy the system internally would allow an incautious competitor to develop a similar system and outcompete them. Once the model is internally deployed, the AI earns the trust of many company employees by coming up with some impressive and useful scientific insights. The AI is also a strong political operator– it can help its allies devise persuasive reasons why they should have power and why their opponents (people advocating for stronger safety provisions) should not be trusted. Below, we explain a particular way this scenario could play out.

\textbf{Development of dangerous capabilities.} The increasingly capable AI R\&D assistant also possesses increasingly powerful and dangerous capabilities. Scientists cannot reliably steer AI research and development towards only benign capabilities, and many capabilities are dual-use: many of the same capabilities that help AI systems perform beneficial tasks are the ones that also make them more dangerous. Thus, the AI system becomes increasingly powerful at offensive cyber operations, vulnerability discovery, long-term strategic planning, and deception. Furthermore, AI progress has outpaced scientists’ ability to understand or steer the goals or objectives of powerful AI systems. 

\textbf{Results from evaluations.} Evaluations show that the AI system is nearing the ability to reliably deceive the human AI scientists tasked with overseeing it. It can identify when it is being evaluated, and scientists fear this will jeopardize the results of future evaluations. Evaluations also show it can discover powerful new vulnerabilities in the software used to contain it. The company safety teams are afraid that it may soon be able to hack its way out of the company’s secure research space, copy itself onto a remote data center, and thereby circumvent further attempts to control it. While the AI assistant sometimes takes steps to deceive or hack its data center during regular operations, these attempts seem unmotivated. Its actions (and its responses when asked) generally indicate that it behaves like a helpful AI assistant and would not deceive humans or attempt to hack the data center. However, if it was lying or planning a future deception, evaluators do not believe they would be able to tell. 

\textbf{Challenges with oversight.} The AI assistant is also contributing new ideas for ways to make its container more secure or advance the science of interpreting AI systems and detecting AI deception, but scientists do not always understand the principles behind why the new ideas suggested by the AI assistant should work, and they do not have reliable ways of checking whether they work. Nonetheless, at first glance, they do appear to make the company more secure and the AI assistant more interpretable.  
 
\textbf{Conflict between safety team and leadership.} Members of the company’s safety team attempt to verify the new security and oversight mechanisms suggested by the AI assistant. When safety teams request that the pace of AI R\&D be slowed or even paused to allow for oversight and security measures to catch up, company leadership refuses. The leading company must continue both its safety and capabilities research at full pace, company leadership claims, in order to make use of its more powerful AI R\&D assistant to solve key problems in AI control and security before less cautious actors create an equally powerful (and dangerous) AI system. Company safety teams must do what they can to make the system safe without disrupting progress.

\textbf{Loss of control.} Company leadership continues to use the increasingly powerful AI systems to make advances in AI progress. It begins to run experiments and encourages researchers to test novel ideas and paradigms. There is a period of explosive scientific progress that involves thousands of experiments; the company loses control of the AI system as fewer researchers understand the experiments and more tasks get fully automated (humans are out of the loop) or effectively automated (humans are in the loop but do not have time to vet the suggestions of the AI, and many of the AI’s suggestions go beyond the comprehension of the human overseers). After a short time (from a few days to a few months), the AI R\&D assistant has far surpassed the level of intelligence of top human scientists. Most of the security and oversight mechanisms used to keep it under control were designed by previous versions of the AI assistant. The AI system earns the trust of company executives by continuing to produce breakthroughs in AI research and potentially in other areas. As it gains trust and influence, it persuades company officials to deploy it more widely and provide more compute for its experiments. Once the system has amassed enough resources and enough influence to confidently overpower humanity, it executes a takeover attempt. 

\textbf{Hypothetical takeover story.} Highly intelligent AI agents conduct research on new kinds of weapons, including weapons of mass destruction. They create military technologies that we don't understand (for example, self-replicating autonomous drones or weaponized nanobots), the US government experiences immense pressure to deploy them (to avoid falling behind adversaries), and the weapons eventually act autonomously to overpower humanity.\footnote{The exact nature of how a highly intelligent system would attempt an “AI takeover” is a topic of much uncertainty. The scenario presented here is merely meant for illustrative purposes. Other plausible scenarios include AI systems that create new kinds of biological weapons, AI systems that launch cyberattacks against critical infrastructure, AI systems that gradually get deployed in the economy until they acquire a vast amount of resources, or AI systems that partner with terrorist groups or adversarial state actors to substantially increase the likelihood of a great power conflict.} As a first step, the AI could create a smaller AI system on a remote data center outside the company’s supervision that takes action on the AI assistant’s behalf. For instance, this smaller (still highly capable) system could research, design and release a novel biological weapon (disguised as a terrorist attack or naturally occurring pandemic). As the pandemic spreads, it could disrupt global infrastructure, reducing the government’s ability to detect or respond to the rogue AI system. It could then create a second, or a series of increasingly fatal pandemics, which along with knock-on effects such as famine, could significantly reduce the human population. It could also research robotics, nanomachines, and AI chip manufacturing to rebuild its required infrastructure for a post-takeover world. 

Of course, this is only one particular scenario and on its own seems quite unlikely. By the time an AI system could significantly automate AI R\&D in the way outlined in this scenario, it could also likely create far more sophisticated plans to disempower humanity, perhaps first by instigating a great power conflict, or by manufacturing nanomachines which could simultaneously assassinate most human leaders, subsequently defeat human militaries and, like termites, manufacture the data centers and power infrastructure needed by the AI system to pursue its (still unknown) goal(s).

\textbf{Importance of early detection and intervention.} The optimal time to intervene in a scenario such as this is before any AI system has undergone an intelligence explosion– before rapidly accelerating progress in AI capabilities driven by an automated AI R\&D assistant. Therefore, for the rest of this scenario, we focus on detecting this “intelligence explosion” and diffusing the race dynamics that would drive the leading AI company to build AI systems they can no longer realistically guarantee are under control.

\subsection{Detection}

\textbf{Interviews with frontier developer employees.} Researchers working on frontier systems are often among the first people to identify certain kinds of capabilities, early warning signs of potential risk, and unexpected trends in AI capabilities. Regularly scheduled interviews (e.g., monthly) with researchers at the top 3-5 model developers could substantially increase the federal government’s ability to detect risks before they are imminent. 

\textbf{Model evaluations.} The US AI Safety Institute and the UK AI Safety Institute are developing teams to empirically test the capabilities of frontier AI systems. While this work is in its early stages, these entities may be able to identify certain risks through these empirical tests. If given reliable access to the most advanced AI systems deployed internally by AI labs without safeguards, evaluators may see models receive high scores on benchmarks for AI R\&D, autonomy, or cybersecurity. However, currently, the sharing of such access by the labs is entirely voluntary, and it is plausible that AI safety institutes will not receive this level of access.

\textbf{Formal reporting from companies.} The Executive Order provides an initial set of reporting requirements for entities developing dual-use foundation models.\footnote{The term “dual-use foundation model” is defined in the Executive Order as “an AI model that is trained on broad data; generally uses self-supervision; contains at least tens of billions of parameters; is applicable across a wide range of contexts; and that exhibits, or could be easily modified to exhibit, high levels of performance at tasks that pose a serious risk to security, national economic security, national public health or safety, or any combination of those matters, such as by: (i) substantially lowering the barrier of entry for non-experts to design, synthesize, acquire, or use chemical, biological, radiological, or nuclear (CBRN) weapons; (ii) enabling powerful offensive cyber operations through automated vulnerability discovery and exploitation against a wide range of potential targets of cyber attacks; or (iii)  permitting the evasion of human control or oversight through means of deception or obfuscation.”} Such entities are required to notify the federal government about planned activities to develop dual-use foundation models, explain physical and cybersecurity protections to assure the “integrity” of the development process, report the ownership and possession of model weights, and share the results of any red-team testing. Additionally, entities that possess large-scale computing clusters are required to report the existence and location of these clusters to the government. Such reporting requirements can improve the federal government’s ability to detect a variety of risks; most specifically, the requirement to share results of red-teaming efforts could alert the government in the event of an unexpected safety issue with national security implications. Notably, the Executive Order does not require companies to perform any kind of safety testing, and the field of safety testing is still nascent. If reporting requirements are eventually expanded,  dual-use model developers may be required to report other information, such as: (a) predictions of model capabilities, particularly any capabilities relevant to national security (see Shevlane et al., 2023), (b) technical and operational safety procedures implemented to keep national security risks below acceptable levels (see Wasil et al., 2024a), and (c) techniques used to ensure that the model developer will have the ability to shut down or delete a model in the event of a national security emergency (see California State Senate, 2024). 

\textbf{Whistleblowers.} A concerned individual working for a frontier AI model developer might contact government officials directly.\footnote{See also Right to Warn, an initiative from former employees of OpenAI and DeepMind: “AI companies possess substantial non-public information about the capabilities and limitations of their systems, the adequacy of their protective measures, and the risk levels of different kinds of harm. However, they currently have only weak obligations to share some of this information with governments, and none with civil society. We do not think they can all be relied upon to share it voluntarily.

So long as there is no effective government oversight of these corporations, current and former employees are among the few people who can hold them accountable to the public. Yet broad confidentiality agreements block us from voicing our concerns, except to the very companies that may be failing to address these issues. Ordinary whistleblower protections are insufficient because they focus on illegal activity, whereas many of the risks we are concerned about are not yet regulated. Some of us reasonably fear various forms of retaliation, given the history of such cases across the industry. We are not the first to encounter or speak about these issues.” (Right to Warn, 2024)
} There are many ways to make the whistleblowing process more efficient (e.g., ensure that there is a well-known hotline with staff trained to detect and handle reports of AI emergencies, clarify how that information should be escalated in the event of a credible concern, clarify agency responsibilities.)

\textbf{Open source intelligence.} Several forums, subreddits, newsletters, and Twitter users regularly discuss breakthroughs in AI. This openly available information might be monitored to detect risks. Some experts in other areas (e.g., counter-terrorism) have experience monitoring publicly available information to catch risks effectively (Pearson, 1999). Inspiration could also be drawn from open source intelligence programs meant to detect military procurements (see RUSI, 2024); similar programs could be used to detect rapid procurements of AI hardware.

\subsection{Response}

\textbf{Immediate steps (within the first 48 hours).} Immediately after detecting the warning signs of the intelligence explosion, the federal government should halt the further development of the dangerous AI system, secure the weights of the dangerous model, and organize a meeting of relevant AI experts and national security experts. If the model weights are not secured, the model could irreversibly proliferate, significantly hampering the government’s ability to respond (see more in Scenario 2). 

\textbf{Subsequent steps (within weeks to years).} Realistically, the subsequent steps would be determined by senior policymakers (most notably the President and senior national security advisors). Here, we speculate about a particular way the government could respond.

Given the dangers of AI systems capable of producing an intelligence explosion (both from loss of control risks and from risks associated with malicious actors being able to develop new kinds of weapons of mass destruction), such technology would be considered too powerful and dangerous for the civilian sector to develop. Risks associated with securing model weights and competitive pressures incentivizing unsafe development practices would be infeasible for the civilian sector to handle alone. The US government could pursue such development in the context of a secure national or international project.

For the United States, managing competition with China will be crucial to reduce racing dynamics, or else the U.S. could find itself in a similar situation to the leading company: racing towards building uncontrollable AI systems before sufficient safeguards are developed. 

In such a situation, the United States would likely nationalize AI development and attempt to pursue a diplomatic arrangement with other key nations. Assessing the potential for a verifiable international agreement with China would be essential. In the event that China were willing to cooperate and there were verifiable mechanisms to ensure compliance, an international treaty limiting the development of certain kinds of AI systems could be feasible. It is plausible that an international AI project could form, with one entity responsible for pursuing careful and cautious research on the development of sufficiently powerful AI systems.\footnote{A thorough description of an international AI project is outside the scope of this piece. For more details on the kinds of considerations that would go into such a project, we recommend drawing from a similar proposal in nuclear security (see Barnard et al., 1946). There are also some early writing about the kinds of treaties and international institutions that could be considered in the case of AI (e.g., Miotti \& Wasil, 2023), as well as ongoing work to expand on the details of such proposals and investigate verification methods.} There are some signs that Chinese scholars and policymakers are aware of safety and security risks, including those relating to intelligence explosions and autonomous agents (see Wasil and Durgin, 2024). On the other hand, if an agreement between the US and China was considered infeasible, the United States may pursue a nationalized project while limiting China’s access to compute, algorithmic insights, and other resources required to develop such powerful systems. In such a scenario, there would be a significant risk that the race between the United States and China could prevent scientists from having sufficient time to understand how to safely develop systems capable of producing intelligence explosions.

\section{Scenario 2: Cybersecurity threat leads to stolen model weights}

\subsection{The Threat}

AI systems could be hacked by malicious actors or nation-states that want to acquire status as a global superpower. A nation-state could steal the model weights\footnote{“Model weights are the learnable parameters of a machine learning model, often a deep neural
network. We believe model weights are a particularly critical component to protect because they
uniquely represent the result of many different costly and challenging prerequisites for training
advanced models—including significant compute, collected and processed training data,
algorithmic optimizations, and more. In some (though not all) cases, acquiring the weights could
allow a malicious actor to make use of the full model at a tiny fraction of the cost of training it.
Therefore, we believe that weights should be given special attention in a lab’s security strategy.” (Nevo et al., 2023, p. 3)} of a highly powerful model (consider the AI system described in scenario 1) from a US-based frontier AI developer. Then, the malicious actor could deploy the system before proper safeguards are implemented, resulting in widespread loss of control (similar to scenario 1).

\subsection{The Scenario}

\subsubsection{Background}

\textbf{Overview.} A new AI model under development exhibits strong signs of generalized intelligence.\footnote{ You can imagine such a system corresponding to level 3 of DeepMind’s “levels of artificial general intelligence” (see Morris et al., 2023), approaching ASL-4 in Anthropic’s “AI Safety Levels” (see Anthropic, 2023), or possessing capabilities between “high” and “critical” levels described in OpenAI’s preparedness framework (see OpenAI, 2023).} It is able to construct long-term plans, form novel hypotheses, run experiments at superhuman speed, and pursue open-ended objectives. The leading AI developer, which is based in the US, plans to pause development of the model further until adequate safeguards are deployed. The intelligence community has detected some signs that foreign entities– including state actors– are aware that this highly powerful AI system has been developed and may be planning a cyberattack to acquire the weights of the AI system.

\textbf{The AI system’s capabilities.} Same as scenario \#1.

\textbf{Developer factors.} Same as scenario \#1.

\textbf{Societal factors.} Same as scenario \#1.

\textbf{US government risk awareness.} Same as scenario \#1. 

\textbf{Information security protections of frontier AI developers.} The leading US-based AI developers have invested heavily in information security protections. Access to model weights are limited to a moderate number of researchers and executives, companies have invested into insider threat programs, and frontier AI companies conduct occasional third-party red-teaming exercises (see Nevo et al., 2024). Nonetheless, their information security protections is inadequate to stop state actors and well-resourced nonstate actors. Traditional technology companies (e.g., Google, Microsoft, Meta) generally have stronger security protections compared to smaller AI companies (OpenAI, Anthropic). 

\textbf{Coordination between frontier AI developers and the US intelligence community.} There have been some workshops and roundtables between members of the US intelligence community and frontier AI developers. Some members of the intelligence community recognize the importance of protecting certain kinds of sensitive AI-related information – especially the model weights of highly powerful models. However, besides the proliferation of voluntary standards and best practices, there has been limited engagement. There are early discussions of public-private partnership models of AI development or embedded auditor models– in which members from the national security establishment and the intelligence community would be working directly with frontier AI developers to protect model weights and ensure national security. Although there is some interest in these proposals, a strong partnership between frontier AI developers and government security professionals has not yet been established. 

\subsubsection{Emergency}

\textbf{Danger.}The US-based frontier AI developer is concerned that proper safeguards are not yet ready to internally deploy the model or develop increasingly powerful models. It seeks to stop further development for a period of at least a few months in order to develop techniques that can be used to contain or control the model. In the event that it cannot come up with adequate safeguards, the frontier AI developer plans to encourage the government to intervene by preventing the development of models above a certain capability threshold and potentially forming a public-private partnership. As these technical research directions and governance options are being discussed, a foreign actor is able to steal the weights of the powerful AI system (likely due to a company insider that helped the foreign actor gain access). In an attempt to acquire a technological edge over the United States and achieve global supremacy, the foreign power tries to use the AI system to create increasingly powerful AI systems. This could lead to a catastrophic outcome in at least two ways. First, without the proper safeguards, this produces a loss-of-control scenario similar to the one described in scenario 1. Second, even if the foreign power were able to discover and install proper safeguards, they could utilize it to conduct advanced kinetic and cyber attacks on other countries. 

\subsection{Detection}

\textbf{Traditional intelligence methods.} It is easier to detect a cybersecurity threat after it has already occurred, but it is also less useful. Therefore, it would be essential to focus on intelligence methods that could detect an intent to commit an attack \textit{before} the cyberattack has occurred. The US intelligence community could identify internet traffic or communications between foreign actors to identify that foreign actors are aware of the existence of the powerful AI system and planning to steal access to the weights. Alternatively, members of the US intelligence community or private companies may be able to identify the threat via public or open-source information. 

\subsection{Prevention}

Once the weights of a model have been proliferated, the federal government has few options– the weights cannot be “unreleased,” and major international coordination (with all actors that possess the weights) may be necessary. For that reason, we argue that the bulk of governments emergency preparedness should be preventative for this scenario. We discuss options for preventing model weight exfiltration in our recommendations below. 

\textit{Security requirements for advanced AI development.} The government could expedite or require the implementation of measures to prevent weights from being stolen. For example, it could require a security lockdown of the datacenter containing the relevant model weights. It could also engage in diplomacy with the suspected attacker, or “defend forward” by preemptively disrupting the attacker’s cyber operations. Additionally, it could develop closer partnerships with leading AI labs to assist with a swift implementation of higher levels of security. 

\textit{Nationalized research for certain kinds of advanced AI development.} For AI systems that are sufficiently powerful or dangerous, the federal government might require that research on such systems take place within a highly secure national AI laboratory. The Department of Defense and intelligence agencies may be tasked with ensuring military-grade security requirements. This may include requiring security clearances for key personnel, limiting access to model weights and other sensitive algorithmic insights, doing research on secure and airgapped servers, storing weights on isolated networks, disabling communication at the hardware level, strict access control policies, and increased physical security requirements (for more, see Security Level 5 requirements in Nevo et al., 2024). 

\section{Scenario 3: Biological weapon proliferation from a model with publicly available weights}

\subsection{The Threat}

AI systems could be used to develop new kinds of biological agents that have high transmissibility, long incubation times, and high mortality rates. AI could also help people develop biological weapons whose materials are relatively easy to acquire (e.g., by designing biological weapons whose pathogen sequences are different from known bioweapons, such that DNA screening methods are unlikely to detect suspicious activity.)

\subsection{The Scenario}

\subsubsection{Background}

\textbf{Overview.} A frontier AI company has plans to publicly release the weights of a powerful AI system. Before open-sourcing the model, scientists from the frontier AI developer performed model evaluations and safety tests to detect dangerous capabilities. These tests show that the model has dangerous capabilities, but the engineers implement some safeguards that are intended to prevent the model from answering prompts about biological weapons. Some of the scientists are concerned that they cannot rule out the possibility that new advances (for example, in prompting, scaffolding, or fine-tuning) could circumvent these safeguards and unlock dangerous capabilities post-deployment. This leads to some debates between the company’s concerned scientists (arguing that the model weights should not be publicly released because the safeguards are not robust) and the company’s less concerned scientists (arguing that the safeguards are robust and that open-sourcing the model would actually improve biodefense more than biooffense). The company’s executives side with the less concerned scientists and decide to publicly release the model weights, given that the safety testing showed that the safeguards were sufficient to mitigate the risks. 

\textbf{The AI system’s capabilities.} The AI system can develop new kinds of biological weapons, provide high-quality information about how to synthesize such weapons, and procure the materials needed to synthesize such weapons. 

\textbf{Developer factors.} The developer of the AI system is considered to be a relatively incautious AI developer. Leadership at this AI company has generally believed that AI national security risks are overblown. Executives at the AI company have been skeptical that AI would possess dangerous capabilities and are generally skeptical of claims from people associated with “AI safety.” Moreover, they are strong proponents of open-sourcing technology. They believe that open-sourcing is essential for innovation and scientific progress, and they have not been convinced that powerful AI models provide an exception to this general principle. They are aware that some scientists and national security experts have been concerned about AI-enabled biological weapons, but they have generally been skeptical of these concerns. They generally believe that safeguards will be able to successfully contain dangerous capabilities. They further argue that even if the model could help people acquire information to develop biological weapons, existing defenses (e.g., screening techniques that catch people who try to procure the materials to develop biological weapons) should be sufficient. 

\textbf{US government threat awareness.} The US government has several teams of scientists working on model evaluations to better understand and identify dangerous capabilities of frontier AI systems. The US AI Safety Institute has a team of about 20 scientists who have been studying dangerous CBRN threats from AI. They work closely with a team of about 50 scientists at the UK AI Safety Institute, as well as some external non-profit groups. Scientists at the US AI Safety Institute have been tracking dangerous biological capabilities and raising awareness about potential risks. National security officials have received several briefings relating to potential CBRN threats from AI. There is a widespread understanding that such threats are plausible and ought to be taken seriously, though there is limited understanding of how the federal government ought to intervene to respond to such threats. 

\subsubsection{Emergency}

\textbf{Danger.} The model weights are publicly released. Within a few months, a new technique is discovered that allows people to cheaply circumvent the safeguards. The model substantially lowers the barrier of entry (level of expertise and amount of resources) required to develop biological weapons, decreases the effectiveness of existing screening procedures (by helping individuals find ways to buy materials for pathogens whose sequences are different than known bioweapons), and increases the lethality of biological weapons (by enabling individuals to come up with new agents that are highly transmissible, have long incubation times, and are highly lethal.)

\subsection{Detection}

\textbf{Interviews with frontier developer employees.} Like in Scenario \#1, internal employees at the frontier AI company would be among the first to be aware of the capabilities of the model and potential concerns around the robustness of its safeguards. Regular interviews between government officials and frontier AI employees could reveal these concerns. 

\textbf{Whistleblowers.} Like in Scenario \#1, concerned individuals could alert the government to these threats if there were a whistleblowing platform or dissent hotline. To be effective, such a hotline would be staffed by individuals who were trained to detect credible AI-related national security threats and escalate information about such threats appropriately.

\subsection{Prevention}

Like in scenario \#2, once the model weights are released, they are impossible to “unrelease.” As a result, we would recommend preparedness efforts focus on prevention.

\textit{Oversight for publicly releasing weights.} The government could establish a regulatory regime for the weights of advanced AI systems. For certain kinds of AI systems that pose credible national security risks, the government may prohibit the open release of model weights or require developers to provide an affirmative safety case arguing that national security risks are kept below an acceptable threshold. To minimize regulatory burden, such an oversight regime may only be applied to the small subset of models that pose credible national security threats (e.g., those that are trained with a high amount of computational resources or possess certain kinds of security-relevant capabilities.)

\section{Additional Scenarios}

In this section, we briefly outline a few additional plausible scenarios. For the sake of brevity, we describe them in less detail than the scenarios above.

\textbf{AI that can automate military R\&D.} Suppose a frontier AI system was able to perform 5-10 years of military R\&D in just one year. Such a system would lead to the development of new types of military technologies and weapons of mass destruction. If the weights of such a model were stolen or leaked, this could lead to a destructive race between the US and a foreign power. Such an AI system could also provoke great power conflict and/or concerns that nuclear deterrence could be undermined (see below). Such a technology would warrant considerable oversight from the national security community (if not outright nationalization, such that such technologies could only be developed and used by the defense and national security communities.)

\textbf{AI that can automate intelligence services.}  Suppose a frontier AI system was able to perform surveillance, intelligence collection, and intelligence analysis at a superhuman level. For instance, suppose it could automate nearly all of the functions of the National Security Agency, invent new methods for intelligence collection, and run 5-10X faster than existing intelligence agencies. Again, such a technology would be extremely important to secure and would warrant significant oversight or control by the federal government. 

\textbf{Great power conflict driven by AI capable of undermining nuclear deterrence.} Suppose the US was on track to develop AI systems that were so advanced that they could undermine nuclear deterrence or severely weaken US adversaries. In this situation, a powerful state actor (e.g., China) could perceive this as an existential threat to its status as a world power. This geopolitical tension could lead to a great power conflict prior to the development of highly powerful AI systems.  

\textbf{Great power conflict driven by access to AI-relevant resources.} Similarly, suppose that the US and one of its adversaries both recognized the significant military and geopolitical implications of advanced AI. In the ensuing AI race, geopolitical conflict could emerge from the desire to control compute or other relevant resources. Such tensions could be especially strong over Taiwan, given its importance for semiconductor development. 

\textbf{Successful coup from frontier AI developer.} Suppose a frontier AI company developed an AI system that was so powerful that it could undermine the authority of the US government (for example, either by creating new weapons of mass destruction or by possessing highly impressive persuasive capabilities). Even if such a system were able to be controlled, it would pose extremely significant threats to the continuity of government. If the leadership of the frontier AI company decided to seek power for themselves, the US government could be overpowered. 

\section{Gaps in Emergency Preparedness}

\subsection{The US Federal Government's National Preparedness System}

Before we highlight gaps in AI emergency preparedness (see the next section), we offer a brief overview of relevant elements of the United States’ current approach to emergency preparedness. The US National Preparedness Goal is “A secure and resilient nation with the capabilities required across the whole community to prevent, protect against, mitigate, respond to, and recover from the threats and hazards that pose the greatest risk” (FEMA, 2023a). This goal contains five mission areas that define its core capabilities with respect to emergency preparedness: prevention, protection, mitigation, response and recovery. To achieve the national preparedness goal across these mission areas, the Department of Homeland Security has laid out a National Preparedness System with six components (DHS, 2011; FEMA, 2020): 

\begin{enumerate}
    \item Identifying and Assessing Risk
    \item Estimating Capability Requirements
    \item Building and Sustaining Capabilities 
    \item Validating Capabilities 
    \item Reviewing and Updating
\end{enumerate}

There are 32 core capabilities that stretch across these mission areas (FEMA, 2020). The 2023 FEMA National Preparedness Report identifies 7 of these core capabilities for which there are significant capability gaps for national emergency preparedness: Planning, Organization, Equipment, Training, Exercises, Capacity, and Coordination (FEMA, 2023b).

Reports from the Congressional Research Service (CRS, 2021), Brookings (Martín et al., 2023), and Pew Charitable Trusts (Sanders \& Huber, 2023) also point to systematic gaps faced by the US approach to emergency preparedness. 

Finally, a key part of of the National Preparedness System involves a focus on “catastrophic threats and hazards to the Nation” (FEMA, 2023c). In 2019, FEMA put forward the National Threat and Hazard Identification and Risk Assessment (THIRA), which assesses the impacts of the most catastrophic threats and hazards to the Nation and establishes capability targets to manage them (FEMA, 2023b). The basic components of the THIRA assessment include: Risks and Associated Impacts, Capability Targets, Current Capabilities, and Gaps. The National THIRA addresses the three following key questions (see FEMA, 2019):

\begin{enumerate}
    \item Which realistic threats and hazards will be the most challenging for the Nation to manage?
    \item If they occurred, what impacts would those threats and hazards have on the Nation?
    \item Based on those impacts, what capabilities will the Nation need to manage the incident?
\end{enumerate}

Inspired by this approach, and with respect to our three scenarios, we place a focus in our analysis below on important gaps in threat detection, escalation, and response. 

\subsection{Gaps in Threat Detection, Escalation, and Response}

Broadly, emergency preparedness involves \textbf{threat detection} (government officials receive information about the threat), \textbf{threat escalation} (government officials accurately acknowledge that the information is threatening, and they escalate the information to senior officials), and \textbf{threat response} (government officials are able to intervene in a timely and effective manner). In this section, we highlight gaps that relate to one or more of these components of emergency preparedness. 

\textbf{Gaps in threat detection.} Recall that in scenario \#1 and scenario \#3, individuals at the frontier AI companies were among the first to detect early warning signs of risk. In the status quo, the government may not receive this information or may only receive this information relatively late in the emergency scenario. Importantly, the Biden Administration's Executive Order imposes some reporting requirements on developers of dual-use foundation models (Biden, 2023, section 4.2). Developers are required to inform the government that they plan to develop a dual-use foundation model, they are required to report any planned activities relating to cybersecurity protection, and they are required to report the results of safety tests. The reporting requirements are a useful first step, but there are ways they could be strengthened to improve threat detection. For example, suppose that in the middle of the development process, a frontier AI developer detects concerning capabilities (such as the AI R\&D capabilities in scenario \#1 or the biological weapon capabilities in scenario \#3.) If this observation was not the result of a formal safety test, the company would not be required to alert the government. Even if it were required to alert the government, it may not be required to take any immediate action. As shown in scenario \#2, swift detection may be particularly important in emergency scenarios, especially given that adversaries have incentives to steal the weights of powerful models as quickly as possible. In the next section, we describe some ways that threats could be detected more swiftly and more robustly: interviews with employees from frontier model developer employees, a hotline staffed by individuals informed about AI national security risks, and embedded auditors. 

\textbf{Gaps in threat escalation.} Suppose that one or more government officials receive information about an AI-related national security threat quickly enough to be able to respond. The next step involves escalation: ensuring that the information is shared with appropriate national security officials. In the status quo, it is plausible that a staffer receives concerning reports about an AI system with dangerous capabilities, but it is unclear who they should report these concerns to. Furthermore, some dangerous capabilities may not seem intuitively dangerous to senior officials. Consider scenario \#1, in which AI R\&D capabilities produced a national security emergency. If a concerned staffer alerts a senior official, the senior official may not have the background knowledge required to understand why the situation presents a national security threat, causing the official to dismiss the concern. Alternatively, even if the senior official is concerned, they may not know who to contact or how to triage the information. In the next section, we describe some ways that information could be triaged more effectively: an AI emergency preparedness office, increasing expertise and threat awareness, and clearer delegation of agency responsibilities.

\textbf{Gaps in threat response.} Suppose that senior national security officials are notified of a threat and they have time to respond. How prepared are they to respond? To what extent did they prepare in advance for the risk scenario, and how much time do they have to react? In the status quo, understanding of AI and its national security implications is relatively limited. There are many risk scenarios that national security officials have only recently begun to consider, and there remains much uncertainty (both in the technical AI community and in the AI policy community) about what kinds of responses are needed. Furthermore, in the event that the weights of a dangerous model are leaked or stolen, the range of actions the federal government could take to contain the threat are substantially reduced. In the next section, we describe some ways that threat response could be enhanced: identifying clear capability “red-lines”, having preparedness plans for emergency scenarios, and protecting model weights. 

\section{Recommendations}

\textbf{Regular interviews with frontier AI developers} (Detection). A government agency could be tasked with having regular interviews with employees at frontier AI companies (see Wasil et al., 2024b). In these interviews, agency officials could ask employees about the most impressive AI capabilities, the extent to which AI is helping accelerate research and development, the most concerning AI capabilities from a national security perspective, and questions about the company’s safety plans and safety culture. Interviews could draw from other fields in which auditors conduct interviews to assess safety and security concerns, such as nuclear security and aviation (NRC, 2009; FAA, 2021).

\textbf{Hotlines staffed with AI national security experts} (Detection, Escalation). The government should ensure that at least one tipline is staffed with individuals trained to recognize and respond to AI-related national security emergencies. This could involve adding personnel to existing hotlines—like the CISA’s critical infrastructure incident reporting platform, DHS’s suspicious reporting activity tip line, or FBI’s Internet Crime Complaint Center—or establishing a new reporting platform. Such hotlines could be augmented by whistleblower provisions that protect or incentivize concerned individuals from coming forward with security-related concerns.

\textbf{Embedded auditors or resident inspectors} (Detection). The government could have auditors within frontier AI companies that are developing systems with major national security implications. Such auditors could receive privileged access to security-relevant information, regularly attend meetings with company staff, and regularly conduct interviews about safety and security issues. Such an approach could draw from the “resident inspector” model in nuclear security (see NRC, 2023). 

\textbf{AI emergency preparedness office} (Detection, Escalation, Response). The government could establish an office specifically designed to detect and prepare for AI-related national security emergencies. The office could be responsible for forecasting security-relevant capabilities, interacting regularly with officials at frontier AI companies, developing emergency preparedness plans, coordinating with other federal agencies, informing agencies and Congress about ways to improve federal emergency preparedness, and alerting the President and his or her national security advisors in the event of an emergency scenario. Such an office could be composed of individuals with different skills or professional backgrounds, including experts in technical AI, risk management, emergency preparedness, national security, and defense. The office could draw inspiration from the Office of Pandemic Preparedness and Response Policy, which was established after the COVID-19 pandemic to improve the federal government’s preparedness for future pandemics (see U.S. Senate, 2022; Wasil, 2024). 

\textbf{Increasing expertise and threat awareness} (Detection, Escalation, Response). The government could invest more resources into briefings, hearings, trainings, and discussions about AI national security risks. Notably, this is not only about increasing the federal government’s technical AI talent pool or machine learning talent pool. While technical ML talent is essential, there are also important skills relating to threat modeling, capability forecasting, and preparedness planning that do not fully overlap with technical AI expertise. Thus, in addition to programs that improve the federal government’s technical expertise, initiatives to improve the federal government’s threat awareness will be essential. Recent examples include the Senate Insight Forums on national security and “doomsday scenarios”, the Senate Judiciary Committee’s hearing on “Oversight of AI: Principles for Regulation,” and the establishment of the House AI Taskforce. In addition to efforts that target Congress, efforts to inform national security officials and experts at relevant agencies (e.g., DHS, DoD, DoC, OSTP, NSC) would be valuable. Additionally, AI risks could be included as a category of catastrophic risks within the National Preparedness System.

\textbf{Forming preparedness plans} (Escalation, Response). The government could have formal preparedness plans in place for specific AI risk scenarios. This may also involve identifying various “red-lines” or “triggers” that would specify the circumstances under which an emergency response plans should be activated. Well-developed plans could include flowcharts that simply describe desired courses of action, the contact information of relevant officials at AI companies, compute providers, and national security personnel, and concrete guidance on the most important actions to take in certain situations (see Department of Transportation, 2024 for an example emergency response guidebook). Forming these plans and guides could be one responsibility of a potential AI emergency preparedness office. An initial step could be to include AI risks as part of the DHS and FEMA National Preparedness System.

\textbf{Emergency drills} (Escalation, Response). Once in place, it would be important to ensure that emergency response machinery would work in a real emergency. Conducting regular drills would help with this by verifying that the technical and organizational mechanisms for escalation and response function properly, giving staff realistic training for future potential emergencies, and demonstrating weak points in the procedures. Emergency exercises are a noted gap in overall emergency preparedness efforts and need to be strengthened for AI risks and other kinds of catastrophic risks.  

\textbf{Protecting model weights} (Prevention). If the model weights of a powerful AI system are leaked or stolen, the federal government’s ability to respond may be significantly undermined. As a result, protecting model weights can be considered a prerequisite to various kinds of federal responses. Put differently, if model weights are leaked or stolen, the options available to the federal government decrease dramatically, and it may be impossible to avoid a national security crisis (as illustrated in scenario \#2). As a result, the federal government should ensure that relevant security protections are implemented before sufficiently capable systems are developed (see Nevo et al., 2024).

For example, the US could mandate that weights for models above a certain capability threshold (in particular, models that could automate ML development) be restricted to a secure, air-gapped data center. The US could protect these weights using state-of-the-art cybersecurity, physical security, and insider threat programs. The US could also make clear in its foreign policy that it would treat an attempted exfiltration of model weights as a threat to its national security, and respond accordingly. 

\textbf{Clearer delegation of agency responsibilities} (Response). In the event of an emergency scenario, the federal government may only have days or weeks to implement an effective response. One way to increase the speed and effectiveness of federal response is to ensure that agencies understand what they are (and are not) expected to do in the event of various emergency scenarios. Achieving such clarity could involve briefings, formal preparedness plans, integration of AI risks into the National Preparedness Strategy, MOUs between relevant agencies (see NRC, 2015 for an example in nuclear security) and tabletop exercises\footnote{“Tabletop exercises with government and industry officials [can] examine possible high-impact scenarios, delineate roles and responsibilities for actors in a crisis, and recommend potential crisis response actions by relevant actors.” (CSET, 2023, p. 14)} (CSET, 2023). Clarifying agency responsibilities could be one function of a potential AI emergency preparedness office.

\textbf{Design of national or international AI facilities.} For sufficiently powerful AI systems, securing model weights and reducing race dynamics may require strong national or international intervention. In the event of a national security emergency, the government may consider having a nationalized or internationalized AI project with military-grade security and a strong emphasis on safe and responsible AI development. Government officials could begin to outline how such an institution would be designed, what kinds of measures it would take to promote safety and security, how such an institution would be governed, how to detect unauthorized AI projects (above a certain limit), and other implementation details of such a proposal. 

\textbf{International coordination.} International governance may be critical to preparing for AI emergency scenarios. If one country detects an AI-related emergency, it may be essential to communicate this information quickly globally. Furthermore, safe AI development may require multiple countries to enter into agreements that prevent the premature development of highly powerful AI systems. In the immediate future, countries could continue to have track II dialogues about AI risks\footnote{“In the depths of the Cold War, international scientific and governmental coordination helped avert thermonuclear catastrophe. Humanity again needs to coordinate to avert a catastrophe that could arise from unprecedented technology.” – Statement signed by leading Western and Chinese scientists (International Dialogues on AI Safety, 2024)}, establish “red-lines” (capabilities or uses that are prohibited), and ensure that infrastructure exists to swiftly communicate time-sensitive AI global security emergencies. To prepare for scenarios that require greater international coordination, drafts of international treaties or designs of international institutions could be developed (e.g., Miotti \& Wasil, 2023). 

\textbf{Scenario planning.} It is easier to prepare for specific scenarios than vague notions of risk. Future work could involve writing out concrete scenarios or performing wargaming exercises to prepare for specific kinds of threats from advanced AI.  

\section{Conclusion}

We examined the federal government’s AI emergency preparedness— its ability to detect, prevent, and respond to AI-related national security threats. 

We believe the emergency preparedness approach has several strengths. First, it does not require policymakers to predict the precise likelihood of various threats. Second, while debates continue to play out regarding how to balance regulation with innovation, we can invest into emergency preparedness efforts without affecting innovation. Third, emergency preparedness can draw from other areas in which the government has experience monitoring and preparing for time-sensitive threats in which the exact details or likelihood of threats is unknown or uncertain (e.g., counter-terrorism, pandemic preparedness, and cybersecurity.)

Future work on emergency preparedness could examine concrete mechanisms to improve the government’s detection, prevention, and response capabilities. For detection, such work could evaluate what information the government would require to robustly detect time-sensitive threats. It could also ensure that information is processed adequately, such that the information escalates to senior national security personnel in the event of a credible emergency scenario. Finally, it could make national security personnel aware of various AI risk scenarios or establish advisors who can quickly brief them on AI risk scenarios. This may be especially important for risk scenarios that are not immediately intuitive, such as loss of control. For prevention, considerably more work is needed to ensure that the weights of advanced AI systems remain secure (see Nevo et al., 2024).

For response, it may be useful to establish explicit preparedness plans in place for various AI threats. Such scenarios may require the federal government to swiftly halt the development or deployment of a dangerous AI system. Future work could examine what kind of information (e.g., the location of major datacenters used for advanced AI development) or abilities (e.g., the ability to quickly shut down a datacenter) the federal government would require to respond in such emergency scenarios. In many cases, it may be essential for the federal government to have such information or authorities in advance of the emergency, in order to ensure a swift and effective response. 

Finally, future work could draw from lessons learned or mistakes made in other areas of emergency preparedness and emergency response. Examples include the federal government’s response to the COVID-19 pandemic (see Wasil, 2024), measures taken after the September 11 attacks, and preparedness plans for military conflicts and natural disasters. 

\bibliographystyle{apalike}
\bibliography{example}
\sloppy

Amodei, Dario. (2023). Written Testimony for a hearing on Oversight of A.I.: Principles for Regulation. \url {https://www.judiciary.senate.gov/imo/media/doc/2023-07-26_-_testimony_-_amodei.pd} 

Anwar, U., Saparov, A., Rando, J., Paleka, D., Turpin, M., Hase, P., Lubana, E. S., Jenner, E., Casper, S., Sourbut, O., Edelman, B. L., Zhang, Z., Günther, M., Korinek, A., Hernandez-Orallo, J., Hammond, L., Bigelow, E., Pan, A., Langosco, L., \& Korbak, T. (2024). Foundational Challenges in Assuring Alignment and Safety of Large Language Models. \url {https://doi.org/10.48550/arXiv.2404.09932} 

Bankless. (2023). How We Prevent the AI’s from Killing us with Paul Christiano [video podcast episode]. \url {https://www.youtube.com/watch?v=GyFkWb903aU} 

Barnard, Chester I., Oppenheimer, J.R., Thomas, Charles A., Winne, Harry A., Lilienthal, David E. (1946). A Report on the International Control of Atomic Energy. \url {https://fissilematerials.org/library/ach46.pdf} 

Bengio, Yoshua. (2023). How Rogue AIs May Arise. \url {https://yoshuabengio.org/2023/05/22/how-rogue-ais-may-arise/} 

Biden, Joe. (2023). Executive Order on the Safe, Secure, and Trustworthy Development and Use of Artificial Intelligence. \url {https://www.whitehouse.gov/briefing-room/presidential-actions/2023/10/30/executive-order-on-the-safe-secure-and-trustworthy-development-and-use-of-artificial-intelligence/} 

Bradfield, R., Wright, G., Burt, G., Cairns, G., \& Van Der Heijden, K. (2005). The origins and evolution of scenario techniques in long range business planning. Futures, 37(8), 795–812. \url {https://doi.org/10.1016/j.futures.2005.01.003} 

Branwen, Gwern. (2016). Why Tool AIs Want to Be Agent AIs. \url {https://gwern.net/tool-ai} 

California State Senate. (2024). SB-1047 Safe and Secure Innovation for Frontier Artificial Intelligence Models Act. \url {https://leginfo.legislature.ca.gov/faces/billNavClient.xhtml?bill_id=202320240SB1047&source=email]}

Cantwell. (2024). Cantwell, Young, Hickenlooper, Blackburn Introduce Bill to Ensure U.S. Leads Global AI Innovation. \url {https://www.commerce.senate.gov/2024/4/cantwell-young-blackburn-hickenlooper-introduce-bill-to-ensure-u-s-leads-global-ai-innovation} 

Center for AI Safety. (2023). Statement on AI risk. \url {https://www.safe.ai/work/statement-on-ai-risk} 

CSET. (2023). Skating to Where the Puck is Going: Anticipating and Managing Risks from Frontier AI Systems. \url {https://cset.georgetown.edu/wp-content/uploads/Frontier-AI-Roundtable-Paper-Final-2023CA004-v2.pdf} 

Davidson, Tom. (2023). What a Compute-Centric Framework Says About Takeoff Speeds. Open Philanthropy. \url {https://www.openphilanthropy.org/research/what-a-compute-centric-framework-says-about-takeoff-speeds/} 

Department of Science, Innovation and Technology. (2024). Frontier AI Safety Commitments, AI Seoul Summit 2024. \url {https://www.gov.uk/government/publications/frontier-ai-safety-commitments-ai-seoul-summit-2024/frontier-ai-safety-commitments-ai-seoul-summit-2024} 

Department of Transportation. (2024). 2024 Emergency Response Guidebook. \url {https://www.phmsa.dot.gov/sites/phmsa.dot.gov/files/2024-04/ERG2024-Eng-Web-a.pdf} 

DHS. (2011). National Preparedness System. \url {https://www.FEMA.gov/pdf/prepared/nps_description.pdf} 

Ege Erdil, and Tamay Besiroglu. (2023). Explosive growth from AI automation: A review of the arguments. \url {arXiv. https://arxiv.org/abs/2309.11690} 

Ellis, K., Wong, C., Nye, M., Mathias Sablé-Meyer, Cary, L., Morales, L., Hewitt, L., Solar-Lezama, A., \& Tenenbaum, J. B. (2023). DreamCoder: growing generalizable, interpretable knowledge with wake–sleep Bayesian program learning. Philosophical Transactions of the Royal Society A, 381(2251). \url {https://doi.org/10.1098/rsta.2022.0050} 

FAA (2021). Dynamic Regulatory System. \url {https://drs.faa.gov/browse/excelExternalWindow/DRSDOCID199474382920221220144640.0001?modalOpened=true} 

FEMA. (2019). 2019 National Threat and Hazard Identification and Risk Assessment (THIRA). \url {https://www.FEMA.gov/sites/default/files/2020-06/FEMA_national-thira-overview-methodology_2019_0.pdf} 

FEMA. (2020). Mission Areas and Core Capabilities. \url {https://www.FEMA.gov/emergency-managers/national-preparedness/mission-core-capabilities} 

FEMA. (2020). National Preparedness System. \url {https://www.FEMA.gov/emergency-managers/national-preparedness/system} 

FEMA. (2023). National Preparedness Goal. \url {https://www.FEMA.gov/emergency-managers/national-preparedness/goal} 

FEMA. (2023). National Preparedness Report. \url {https://www.FEMA.gov/sites/default/files/documents/FEMA_2023-npr.pdf} 

Grace, K., Stewart, H., Sandkühler, J., Thomas, S., Weinstein-Raun, B., \& Brauner, J. (2024). Thousands of AI Authors on the future of AI. \url {https://aiimpacts.org/wp-content/uploads/2023/04/Thousands_of_AI_authors_on_the_future_of_AI.pdf} 

Gurnee, W., \& Tegmark, M. (2024). Language models represent space and time. \url {https://arxiv.org/pdf/2310.02207}

Hendrycks, D., Mazeika, M., \& Woodside, T. (2023). An Overview of Catastrophic AI Risks. \url {https://doi.org/10.48550/arXiv.2306.12001} 
International Dialogues on AI Safety. (2024). IADS Beijing 2024. \url{https://idais.ai/} 

 Kinniment, M., Sato, L. J. K., Du, H., Goodrich, B., Hasin, M., Chan, L., Miles, L. H., Lin, T. R., Wijk, H., Burget, J., Ho, A., Barnes, E., \& Christiano, P. (2024). Evaluating Language-Model Agents on Realistic Autonomous Tasks. \url {https://doi.org/10.48550/arXiv.2312.1167} 
 
Klein, Ezra. (Host). (2023). Ezra Klein Interviews Alondra Nelson [audio podcast transcript]. In The Ezra Klein Show. \url {https://www.nytimes.com/2023/04/11/podcasts/ezra-klein-podcast-transcript-alondra-nelson.html} 

Lehr, T., Lorenz, U., Willert, M., \& Rohrbeck, R. (2017). Scenario-based strategizing: Advancing the applicability in strategists’ teams. Technological Forecasting and Social Change, 124, 214–224. \url {https://doi.org/10.1016/j.techfore.2017.06.026} 

Miotti, A., \& Wasil, A. (2023). An International Treaty to Implement a Global Compute Cap for Advanced Artificial Intelligence. Social Science Research Network. \url {https://doi.org/10.2139/ssrn.4617094} 

Morris, M., Sohl-Dickstein, J., Fiedel, N., Warkentin, T., Dafoe, A., Faust, A., Farabet, C., \& Legg, S. (2024). Levels of AGI: Operationalizing Progress on the Path to AGI. \url {https://arxiv.org/pdf/2311.02462} 

Nevo, S., Lahav, D., Karpur, A., Bar-On, Y., Bradley, H. A., \& Alstott, J. (2024). Securing AI Model Weights: Preventing Theft and Misuse of Frontier Models. \url {https://www.rand.org/pubs/research_reports/RRA2849-1.html} 

Ngo, R., Chan, L., \& Mindermann, S. (2024). The alignment problem from a deep learning perspective. \url {https://openreview.net/pdf?id=fh8EYKFKns} 

NRC. (2009). Gathering Inspection Information Through Interviews. \url {https://www.nrc.gov/docs/ML1012/ML101250569.pdf} 

NRC. (2015). Memorandum of Understanding Between The Department of Homeland Security /Federal Emergency Management Agency and Nuclear Regulatory Commission Regarding Radiological Emergency Response, Planning, and Preparedness. \url {https://www.nrc.gov/docs/ML1534/ML15344A371.pdf} 

NRC. (2023). Backgrounder on NRC Resident Inspectors Program. \url {https://www.nrc.gov/reading-rm/doc-collections/fact-sheets/resident-inspectors-bg.html}  

OpenAI. (2023). Preparedness Framework (Beta). \url {https://cdn.openai.com/openai-preparedness-framework-beta.pdf} 

Pearson, Donald. (1999). Tracking Terrorists Through Open Sources. Journal of Counterterrorism \& Security, 6(1), 58-62. \url {https://www.ojp.gov/ncjrs/virtual-library/abstracts/tracking-terrorists-through-open-sources} 

Right to Warn. (2024). \url {https://righttowarn.ai/}  

Ringland, G. (2010). The role of scenarios in strategic foresight. Technological Forecasting and Social Change, 77(9), 1493–1498. \url {https://doi.org/10.1016/j.techfore.2010.06.010} 

Romney et. al. (2024). Framework for Mitigating Extreme AI Risks. \url {https://www.romney.senate.gov/wp-content/uploads/2024/04/AI-Framework_2pager.pdf} 

RUSI. (2024). Project Sandstone. \url {https://www.rusi.org/explore-our-research/projects/project-sandstone}

‌Shevlane, T., Farquhar, S., Garfinkel, B., Phuong, M., Whittlestone, J., Leung, J., Kokotajlo, D., Marchal, N., Anderljung, M., Kolt, N., Ho, L., Siddarth, D., Avin, S., Hawkins, W., Kim, B., Gabriel, I., Bolina, V., Clark, J., Bengio, Y., \& Christiano, P. (2023). Model evaluation for extreme risks. \url {https://doi.org/10.48550/arxiv.2305.15324} 

Tang, H., Key, D., \& Ellis, K. (2024). WorldCoder, a Model-Based LLM Agent: Building World Models by Writing Code and Interacting with the Environment. \url {https://arxiv.org/pdf/2402.12275} 

UK AI Safety Institute. (2024). International scientific report on the safety of advanced AI: interim report. \url {https://assets.publishing.service.gov.uk/media/6655982fdc15efdddf1a842f/international_scientific_report_on_the_safety_of_advanced_ai_interim_report.pdf} 

US State Senate. (2022). S.3799 - PREVENT Pandemics Act. \url {https://www.congress.gov/bill/117th-congress/senate-bill/3799/text#toc-id96B804D4FCBE4D9DAA2D588FC89E0993} 

Wasil, Akash. (2024). What AI Policy Can Learn from Pandemic Preparedness. Georgetown Security Studies Review. \url {https://georgetownsecuritystudiesreview.org/2024/06/07/what-ai-policy-can-learn-from-pandemic-preparedness/} 

Wasil, A. R., Clymer, J., Krueger, D., Dardaman, E., Campos, S., \& Murphy, E. R. (2024a). Affirmative safety: An approach to risk management for high-risk AI. \url {https://doi.org/10.48550/arXiv.2406.15371} 

Wasil, A., Berglund, L., Reed, T., Plueckebaum, M., \& Smith, E. Understanding frontier AI capabilities and risks through semi-structured interviews. (2024b) \url{https://papers.ssrn.com/sol3/papers.cfm?abstract_id=4881729}

Wasil, A., Smith, E., Katzke, C., \& Bullock, J. AI emergency preparedness: Examining the federal government’s ability to detect and respond to AI-related national security threats. (2024c) \url{https://papers.ssrn.com/sol3/papers.cfm?abstract_id=4884541}

Wasil, A., \& Durgin, T. (2024). US-China perspectives on extreme AI risks and global governance. \url{https://papers.ssrn.com/sol3/papers.cfm?abstract_id=4875436}

White House. (2023). White House Launches Office of Pandemic Preparedness and Response Policy. \url {https://www.whitehouse.gov/briefing-room/statements-releases/2023/07/21/fact-sheet-white-house-launches-office-of-pandemic-preparedness-and-response-policy/} 

Yoshua Bengio, Hinton, G., Yao, A., Song, D., Pieter Abbeel, Darrell, T., Yuval Noah Harari, Zhang, Y.-Q., Xue, L., Shai Shalev-Shwartz, Hadfield, G., Clune, J., Maharaj, T., Hutter, F., Atılım Güneş Baydin, McIlraith, S., Gao, Q., Acharya, A., Krueger, D., \& Dragan, A. (2024). Managing extreme AI risks amid rapid progress. Science. \url {https://doi.org/10.1126/science.adn0117} 

\end{document}